\documentstyle[pra,eqsecnum,aps]{revtex}
\draft
\begin{document}
\title{Non-classical Photon Statistics For Two-mode Optical Fields}
\author{Arvind\cite{email}}
\address{Department of Physics\\
Indian Institute of Science,  Bangalore - 560 012, India}
\author{N. Mukunda\cite{jncasr}}
\address{Center for Theoretical Studies and Department of Physics\\
Indian Institute of Science,  Bangalore - 560 012, India}
\date{\today}
\maketitle
\begin{abstract}
The non-classical property of subpoissonian photon statistics
is extended from one to two-mode electromagnetic fields,
incorporating the physically motivated property
of invariance under passive unitary transformations.
Applications to squeezed coherent states,
squeezed thermal states, and superposition of coherent states
are given. Dependences of extent of non-classical behaviour on the
independent squeezing parameters are graphically displayed.
\end{abstract}
\section{Introduction}
\label{introduction}
\setcounter{equation}{0}
Non-classical properties and effects of radiation~\cite{nonclassical}
have received considerable attention in the past two decades
and continue to be an active area of research.
Quadrature squeezing~\cite{quadrature-sq},
subpoissonian photon statistics (SPS),
and antibunching~\cite{antibunching} of photons are three
prominent examples of such properties leading to measurable effects.
Quadrature squeezing is related to the reduction of
noise in one of the two quadrature components
below the coherent state value,
and has been both theoretically and experimentally studied
for one-mode as well as multi-mode fields. Antibunching
arises when the photon number distribution becomes
subpoissonian leading to anticorrelation in the photons
detected in a typical detection experiment. In all
these cases, the diagonal coherent state description of the
fields involved does not have a classical interpretation
and hence no classical description can explain these effects.
The extension from one to two or more modes for the case of
quadrature squeezing is nontrivial and leads to new
physical effects~\cite{two-mode}~\cite{multimode}.
The phenomenon of SPS has been formulated~\cite{mandel} and
observed~\cite{exp-SPS} primarily for one-mode
situations. For situations involving two or more modes,
in the existing literature, such properties are
invariably studied for one of the modes or a predefined linear
combination of the modes~\cite{pair-coherent}.
Such an analysis cannot be used to make any clear
statement about the classical
or non-classical nature of the field involved, because the
linear combination of modes which may show SPS
may in general be different from the mode chosen for the
analysis. Another kind of generalisation of
SPS to two-mode fields has been done using
a particular inequality involving the correlation
between the two modes; however this does
not exhaust the possibilities available at the
level of quadratic expressions in photon
number~\cite{SPS-two-correlation}.
This clearly indicates the need for a more
satisfactory way of looking at non-classical statistics,
for fields involving two or more modes.

Our aim in this paper is to develop a notion of SPS for
two-mode fields which is intrinsically two-mode in character,
can be used in an unambiguous way to make a statement about the
classical or non-classical nature of the field, and has
physically reasonable invariance properties. The group of
linear, homogeneous, canonical transformations $Sp(4,\Re)\/$,
the symmetry group basic to the quantum mechanical description
of the two-mode field, naturally splits into two parts: the
photon number conserving (maximal compact) passive subgroup
$U(2)\/$, and the photon number nonconserving (non-compact)
active part. The maximal compact subgroup $U(2)\/$,
while acting on the Hilbert space of the two-mode system
through its unitary representation, is incapable of
generating a non-classical(classical) state starting from a
classical(non-classical) one because the diagonal
coherent state distribution function is
covariant under such transformations.
Therefore, it is reasonable to require that
any signature of non-classicality for a two-mode system,
in particular SPS, be $U(2)\/$
invariant. To achieve this we regard all modes related to the
original ones by passive $U(2)\/$ transformations as basically
equivalent; then a survey of the SPS properties
for each mode in this equivalence class of modes leads to
the formulation of a $U(2)\/$ invariant definition of SPS.
We search over the set of all modes for that one which minimizes
the relevant parameter measuring number fluctuation minus the
mean. In this way, we arrive at that
$U(2)\/$ combination of the two modes which is most
likely to be manifestly subpoissonian.
A much wider class of non-classical states
can be explored using this formalism compared to the
earlier ways of handling two-mode situations.

The material in this paper is arranged as follows:
In Section~\ref{the-definition} we recapitulate the basic
kinematics of two-mode systems and the action of the group
$Sp(4,\Re)\/$ on the nonhermitian annihilation and creation
operators. The hermitian generators of this action, and the
maximal compact subgroup $U(2) \in Sp(4,\Re)\/$, are recorded.
The notion of $U(2)\/$-invariant SPS is then developed
by regarding all modes related to one another by (passive)
$SU(2)\/$ transformations as equivalent, and by minimizing the
variable one-mode $Q$ parameter over the group $SU(2)\/$.
The algebraic machinery needed to carry this out, for an
arbitrary given state of the two-mode  system, is set up.
In Section~\ref{sq-coh} we consider three applications:
squeezed coherent states, squeezed thermal states and a general
superposition of two coherent states.
In each case the analytic work
is carried out as far as possible, and then we resort to
numerical studies which are graphically displayed.
Section~\ref{con-remarks} contains concluding remarks.
\section{$U(2)$ Invariant Definition Of Subpoissonian Photon
Statistics For Two-mode Systems}
\label{the-definition}
\setcounter{equation}{0}
We consider two orthogonal modes of the radiation field,
their orthogonality being achievable by their having different
frequencies, orthogonal polarizations or different directions
of propagation. These modes can be quantum mechanically
described by photon annihilation operators $a_r\/$ and corresponding
photon creation operators $a_r^{\dag}\/$, where $r=1,2$.
These operators can be arranged as a column vector $\xi^{(c)}\/$:
\begin{equation}
\xi^{(c)}  =  (\xi^{(c)}_a) = \left( \begin{array}{c}
a_1 \\ a_2 \\ a^\dagger_1 \\ a^\dagger_2 \end{array}
\right), \quad a=1\cdots4.
\end{equation}
The superscript $(c)\/$ on $\xi\/$ indicates that
the entries here are complex i.e. nonhermitian.
The quadrature components of these operators,
which are the hermitian phase space variables $q$'s and
$p$'s, can be written as another column vector, related
to $\xi^{(c)}\/$ by a fixed numerical matrix $\Omega$:
\begin{equation}
\xi=(\xi_a) = \left(\begin{array}{c} q_1\\ q_2\\ p_1\\ p_2
\end{array} \right) =
\Omega^{-1} \xi^{(c)},\quad
\Omega = (\Omega^{-1})^\dagger =   \frac{1}{\sqrt{2}}
\left(\begin{array}{cccc}
1&0&i&0\\0&1&0&i\\1&0&-i&0\\0&1&0&-i \end{array} \right)
\end{equation}
The canonical commutation relations obeyed by the creation
and annihilation operators can be written in terms of $\xi$
or $\xi^{(c)}$:
\begin{eqnarray}
[\xi_a ,\xi_b ] & = & i \beta_{ab}, \nonumber\\
\mbox{}
[\xi^{(c)}_a ,\xi^{(c)}_b ] & = & \beta_{ab}, \nonumber\\
(\beta_{ab}) & = & \left(\begin{array}{cccc}
0&0&1&0\\0&0&0&1\\-1&0&0&0\\0&-1&0&0 \end{array}\right),
\label{can-com-rel}
\end{eqnarray}

A general real linear homogeneous
transformation on the $q$'s and $p$'s which preserves these
commutation relations is described
by a $4\times 4\/$ real matrix $S\/$ obeying the condition:
\begin{equation}
S\; \beta\; S^T = \beta.
\label{symp-cond}
\end{equation}
This is the defining property for the elements of
the non-compact group $Sp(4,\Re)$:
\begin{equation}
Sp(4,\Re)=\left\{ S=4\times 4\; \mbox{real matrix}\;\Bigm\vert
S\; \beta\; S^T = \beta \right\}.
\end{equation}
When $\xi\/$ undergoes a transformation by $S\in Sp(4,\Re)\/$,
the nonhermitian operators $\xi^{(c)}$ transform through a
complex matrix $S^{(c)}\/$, obtained from $S$ by conjugation
with $\Omega \/$:
\begin{eqnarray}
S\in Sp(4,\Re):&\xi^\prime = S\xi\Rightarrow\nonumber\\
& \xi^{(c)}{}^\prime=S^{(c)}\; \xi^{(c)}, \nonumber \\
& S^{(c)} =\Omega \;S \;\Omega^{\dagger}.
\label{op-trans}
\end{eqnarray}
The complex matrices $S^{(c)}\/$ are a faithful
representation of the real matrix group $Sp(4,\Re)\/$. In this
sense we will treat them as elements of $Sp(4,\Re)$.

The maximal compact subgroup $U(2)\/$ of
$Sp(4,\Re)\/$ can be identified as follows:
\begin{eqnarray}
{\cal K}\equiv U(2)&=&\{S^{(c)}(U)
\in Sp(4,\Re)\vert U\in U(2)\} \subset
Sp(4,\Re), \nonumber \\
S^{(c)}(U)&=&\left( \begin{array}{cc} U&0\\0&U^{\star}
\end{array} \right).
\label{u2-def}
\end{eqnarray}
The block diagonal form is responsible for the fact that
such transformations do not mix $a\/$ and $a^{\dagger}\/$;
in fact ${\cal K}\/$ is the largest subgroup with this property.

Let ${\cal H}\/$ be the Hilbert space on
which $\xi\/$ and $\xi^{(c)}\/$
act irreducibly. It follows
from the Stone-von Neumann theorem~\cite{stone-von} that,
since the canonical commutation and hermiticity
relations are invariant under the
transformation~(\ref{op-trans})
for any $S^{(c)} \in Sp(4,\Re)\/$,
it is possible to construct a unitary operator ${\cal U}(S^{(c)})\/$
on ${\cal H}$ implementing~(\ref{op-trans}) via conjugation:
\begin{eqnarray}
S^{(c)} \in Sp(4,\Re)\/:\/S^{(c)}_{ab}\xi^{(c)}_b =
{\cal U}(S^{(c)})^{-1}\xi^{(c)}_a{\cal U}(S^{(c)}), \nonumber \\
{\cal U}(S^{(c)})^{\dag}{\cal U}(S^{(c)})= {\bf 1}\quad \mbox{on}
\quad {\cal H}.
\label{hspace-action}
\end{eqnarray}
The generators of the operators ${\cal U}(S^{(c)})\/$ are
given by ten independent, hermitian, quadratic expressions in $a_r\/$
and $a^{\dagger}_{r}\/$. We define the four photon number
conserving generators $J_0$, $J_j\/$ and the six photon
number non-conserving generators $K_j$, $L_j$, $j=1,2,3\/$:
\begin{mathletters}
\label{sp4r-gen}
\begin{eqnarray}
J_0 &=&
\frac{1}{2}(N+1)
= \frac{1}{2}(a_1^\dagger a_1 + a_2^\dagger a_2 + 1);\\ J_1 &=&
\frac{1}{2}(a^{\dagger}_1 a_2
+ a^{\dagger}_2 a_1), \nonumber \\ J_2 &=&
\frac{i}{2}(a^{\dagger}_2 a_1 -
a^{\dagger}_1 a_2), \nonumber \\ J_3 &=&
\frac{1}{2} (a^{\dagger}_1a_1 - a^{\dagger}_2a_2); \\
K_1 &=&
\frac{1}{4}(a_1^{\dagger}{}^{2} + a_1^2 -
a_2^{\dagger}{}^2 - a_2^2),\nonumber\\ K_2 &=& -
\frac{i}{4}(a_1^{\dagger}{}^2 - a_1^2 + a_2^{\dagger}{}^2
- a_2^2 ),\nonumber \\
K_3 &=& -\frac{1}{2}(a^{\dagger}_1a^{\dagger}_2 +a_1a_2);\\
L_1 &=& \frac{i}{4}(a_1^{\dagger}{}^2
- a_1^2 - a_2^{\dagger}{}^2 + a_2^2 ), \nonumber \\ L_2 &=&
\frac{1}{4}(a_1^{\dagger}{}^2 + a_1^2 +
a_2^{\dagger}{}^2 + a_2^2 ),\nonumber\\ L_3 &=& -\frac{i}{2}
(a^{\dagger}_1 a^{\dagger}_2 - a_1a_2).
\end{eqnarray}
\end{mathletters}
These generators obey the commutation relations
\begin{mathletters}
\label{comm-rel}
\begin{eqnarray}
\mbox{}[J_j,J_k] & = & i\epsilon_{jkl}J_l,\nonumber\\
\mbox{}[J_{0},J_j] & = & 0;\label{u2-comm}\\
\mbox{}[J_j,K_k\;\mbox{or}\;L_k] & = & i\epsilon_{jkl}(K_l
\;\mbox{or}\;L_l),\nonumber\\
\mbox{}[J_{0},K_j \pm iL_j] & = & \mp (K_j \pm iL_j);
\label{u2-ncmp-comm}\\
\mbox{}[K_j,K_k] & = & [L_j,L_k] = -i\epsilon_{jkl}J_l,\nonumber\\
\mbox{}[K_j,L_k] & = & i\delta_{jk}J_{0}.
\label{ncmp-comm}
\end{eqnarray}
\end{mathletters}
{}From the above commutation relations, it is clear that $J_{0}\/$ and
$J_j\/$ form the algebra of $U(2)\/$ and hence
generate the unitary operators corresponding to the
elements of the maximal compact subgroup ${\cal K}\/$ of
$Sp(4,\Re)\/$. On the other hand, $K_j\/$ and $L_j\/$
are the generators of the unitary operators  corresponding
to the non-compact elements of $Sp(4,\Re)$ and they do not
form a closed algebra. These non-compact elements are actually
the squeezing transformations and their complete classification
has been given elsewhere~\cite{two-mode}.

We now consider the notion of SPS for the physical states
of a two-mode system. For one-mode systems,
such an analysis is simple and is based
on Mandel's $Q\/$ parameter~\cite{mandel}:
\begin{equation}
 Q= \frac{\langle a^{\dag
2}a^{2}\rangle - \langle a^{\dag}
a\rangle^2}{\langle a^{\dag}a\rangle}
\end{equation}
where $a\/$ and $a^{\dag}\/$ are the annihilation and
creation operators for the one-mode radiation field, the
expectation values being taken for the state of interest.
The $Q\/$ parameter  distinguishes between physical
states as having poissonian, subpoissonian and superpoissonian
photon statistics, as
$Q\/$ is $0, <0$ and $>0$ for the above cases respectively.
In particular, the states with negative $Q\/$ are non-classical,
in the sense that such a distribution can not be derived
from any classical statistical ensemble. Therefore, in this
limited sense, the $Q\/$ parameter
can be used to classify states as classical and non-classical.
More precisely, $Q<0\,(>0)\/$ is a sufficient(necessary) condition
for non-classicality(classicality).

For a situation involving two modes, the notion of
SPS defined above
is not appropriate. At the most, one can analyze the
photon statistics of one of the modes, or a preselected
linear combination of both. Then again, for a given state,
this mode which one chooses
need not be the one in which the photon number distribution
may be non-classical. Hence the sign of $Q\/$ for the preselected
mode may not disclose the non-classical nature of the two-mode state,
even if it is non-classical. This clearly
indicates that an intrinsically two-mode notion of
SPS is required.

The standard way~\cite{nonclassical}
of distinguishing the classical from the
non-classical states (already implicitly assumed in the above)
is through the diagonal coherent state description.
The general two-mode coherent state with complex two-component
displacement $\tilde{z} = (z_1 ,z_2)\/$ is defined by
\begin{eqnarray}
\vert\tilde{z}\rangle & = &
\exp\left(\tilde{z}\cdot\tilde{a}^\dagger
- \tilde{z}^\star\cdot\tilde{a}\right)\;\vert 0,0\rangle
\nonumber\\
& = & \exp\left(-\frac{1}{2}\vert z_1\vert^2 - \frac{1}{2}\vert
z_2\vert^2\right)\;\exp\left(z_1 a_1^\dagger +
z_2 a_2^\dagger \right)\;\vert 0,0\rangle.
\label{coh-st-def}
\end{eqnarray}
These are normalized states and form an over-complete set.
A given two-mode density operator $\rho\/$  can be expanded
in terms of them:
\begin{equation}
\rho = \int \frac{d^2z_1 d^2z_2}{\pi^2} \phi(z_1,z_2)
\vert z_1,z_2\rangle \langle z_1,z_2\vert
\label{phi}
\end{equation}
The unique normalized weight function $\phi(z_1,z_2)\/$
gives the complete description of the
two-mode state and can in general be a
distribution which is quite singular~\cite{klouder-sudarshan}.
In the case when $\phi(z_1,z_2)\/$ can be
interpreted as a probability distribution
(i.e. it is nonnegative and is nowhere more singular
than a delta function), equation~(\ref{phi}) implies
that the state $\rho\/$ is a classical mixture
of coherent states which have a natural classical limit.
Such quantum states are referred to as classical;
in contrast the others for which $\phi(z_1,z_2)\/$
either becomes negative or more
singular than a delta function somewhere, are
defined as being non-classical.
This classification is general and can be done for any
number of modes. In particular, for the one-mode case,
the states having negative $Q\/$ are a subset
of the states with non-classical diagonal coherent state
distribution functions.

When the two-mode state, with density matrix
$\rho\/$, transforms under a unitary operator
corresponding to the compact $U(2)\/$ subgroup
of $Sp(4,\Re)\/$, the distribution $\phi(z_1,z_2)\/$
undergoes a point transformation given in terms of the
$U(2)\/$ matrix $U\in U(2)\/$:
\begin{eqnarray}
\rho^{\prime} = {\cal U}(S^{(c)}(U)) \rho {\cal U}(S^{(c)}(U))^{-1}
\Leftrightarrow \phi^{\prime}(z_1,z_2) =
\phi(z_1^{\prime},z_2^{\prime}),
\nonumber \\
\left(\begin{array}{c} z_1^{\prime} \\
z_2^{\prime} \end{array} \right)
= U\left(\begin{array}{c} z_1 \\ z_2 \end{array} \right)
\end{eqnarray}
Thus, under $U(2)\/$ transformations classical states
map on to classical ones and non-classical states
to non-classical ones; these  transformations are incapable of
generating a non-classical state from a classical one.
Therefore, it is reasonable to demand that any signature of
non-classicality be invariant under such transformations.

At this stage, we recapitulate and collect
some interesting and important properties of
the maximal compact subgroup ${\cal K}\/$ of $Sp(4,\Re)\/$:
\begin{itemize}
\item[(a)] As is clear from eqn.(\ref{u2-def}),
when $\xi^{(c)}$ undergoes a $U(2)\/$ transformation, the
annihilation operators $a_r$'s are not mixed with the
creation operators $a_r^{\dag}$'s.
\item[(b)] The action of the elements of $U(2)$ (generated by
$J_{0}$ and $J_j$) on a state does not change the total photon
number or its distribution.
\item[(c)] The diagonal coherent state distribution
function is covariant under $U(2)\/$ transformations.
\item[(d)] One requires only passive optical elements to
experimentally implement any $U(2)\/$ transformation on
a state of the two-mode  electromagnetic field~\cite{u2-exp}.
\end{itemize}

Motivated by the above considerations we now
define an intrinsically two-mode and
$U(2)\/$ invariant notion of SPS.
For the purpose of our present analysis it is
convenient to define $U(2)\/$ transformed mode operators
in terms of two column vectors ${\cal A}\/$ and $\alpha
\/$:
\begin{equation}
{\cal A}=\left( \begin{array}{c}a_1\\ a_2
\end{array} \right), \quad
\alpha=\left( \begin{array}{c}\alpha_1\\ \alpha_2
\end{array} \right)
\end{equation}
where $\alpha_1$ and $\alpha_2$ are complex numbers such that:
\begin{eqnarray}
U(\alpha)&=&\left(\begin{array}{cc}
\alpha_{1}^{\star} & \alpha_{2}^{\star} \\
-\alpha_2 &  \alpha_1
\end{array} \right) \in SU(2), \quad
\vert\alpha_1 \vert^2 + \vert \alpha_2 \vert^2 = 1\,;
\nonumber \\
U(\alpha,\psi)&=&\left(\begin{array}{cc}
\alpha_{1}^{\star} & \alpha_{2}^{\star} \\
-e^{i\psi }\alpha_2 & e^{i\psi } \alpha_1
\end{array} \right) \in U(2),\quad 0\leq \psi \leq 2\pi
\end{eqnarray}
When $\xi^{(c)}\/$ undergoes a $U(2)\/$ transformation
given by $U(\alpha,\psi)\/$, the annihilation and creation operators
for the first transformed mode can be written in
terms of ${\cal A}\/$ and $\alpha\/$ alone:
\begin{eqnarray}
 a(\alpha)&=&\alpha^{\dag}{\cal A}=\alpha_1^{\star}a_1
+\alpha_2^{\star}a_2 \nonumber \\
a(\alpha)^{\dagger}&=&{\cal A}^{\dagger}
\alpha=\alpha_1a_1^{\dagger}
+\alpha_2 a_2^{\dagger}.
\end{eqnarray}
Thus the most general normalized "first mode" after the
$U(2)\/$ transformation is determined by $SU(2) \in U(2)\/$
independent of $\psi\/$.
This particular mode will henceforth be called the $SU(2)\/$
transformed mode, and $\alpha\/$ will be used to denote the
$SU(2)\/$ element involved.

Let $\rho \/$ be the density matrix for any (pure or mixed)
state of the two-mode radiation field.
Then we can define the following function:
\begin{eqnarray}
Q(\rho;\alpha) &=& \frac{\langle a(\alpha)^{\dag
2}a(\alpha)^{2}\rangle_\rho  - \langle a(\alpha)^{\dag}
a(\alpha)\rangle^{2}_\rho}{\langle {\cal A}^{\dag}{\cal A}\rangle_{\rho}}
\nonumber \\
		&=& \frac{Tr ( \rho a(\alpha)^{\dag
2}a(\alpha)^{2})  - (Tr ( \rho a(\alpha)^{\dag}
a(\alpha)))^{2}}{Tr ( \rho {\cal A}^{\dag}{\cal A})}
\end{eqnarray}
which is similar to the Mandel Q parameter for the
$SU(2)\/$ transformed mode $a(\alpha)$.

When the state $\rho \/$ is transformed by the unitary operator
${\cal U}(S^{(c)}(U))\/$ for some $ U\in U(2)\/$, the function
$Q(\rho;\alpha)\/$ can be shown to change covariantly:
\begin{eqnarray}
S^{(c)}(U) \in {\cal K} \;:\; \rho^{\prime}&=&
{\cal U}(S^{(c)}(U))\rho {\cal U}(S^{(c)}(U))^{-1}
\Rightarrow  \nonumber \\
 Q(\rho^{\prime};\alpha)&=&Q(\rho;\alpha^{\prime}),\quad
\alpha^{\prime}=U \alpha
\end{eqnarray}
Now an overall phase change corresponding to elements in the
$U(1)\/$ subgroup of $U(2)\/$ actually leaves $Q(\rho;\alpha)\/$
unchanged, therefore no dependence on $\psi\/$ has been shown.
So we have the freedom of
running over all $\alpha$'s $\in SU(2)\/$
i.e. we can choose various
linear combinations of the two modes involved, related to
each other by $SU(2)\/$ transformations. Since we want to
unearth the signature of the non-classical
nature (if present) of the photon statistics,
we vary $\alpha\/$ till we reach
the minimum value of the function $Q(\rho;\alpha)\/$:
\begin{equation}
Q(\rho)=
\mathop{\mbox{Min}}\limits_{{\mbox{over all}}\atop{\alpha \in SU(2)}}
Q(\rho;\alpha)=Q(\rho;\overline{\alpha})\quad
\mbox{s.t.} \quad Q(\rho;\overline{\alpha}) \leq Q(\rho;\alpha)
\end{equation}
If $Q(\rho) < 0\/$ we conclude that the
photon number distribution for the two-mode state $\rho \/$
has a non-classical feature and we call it
subpoissonian, or amplitude squeezed. This is our $U(2)\/$
invariant definition of SPS for states of two-mode
fields.  The mode in which
the subpoissonian nature is manifest to the maximum degree
is $a(\overline{\alpha})$.

The numerator in our definition of $Q(\rho;\alpha)\/$
consists of two terms, one arising from the
expectation values of quadratic expressions in the creation and
annihilation operators and the other arising from the expectation
values of quartic terms. The quadratic  term can be
written:
\begin{eqnarray}
Tr(\rho a(\alpha)^{\dagger} a(\alpha))=
s + \tilde{u}.\tilde{q},\nonumber\\
\tilde{q}=\tilde{q}(\alpha)=
\alpha^{\dagger}\tilde{\sigma}\alpha,
\end{eqnarray}
with the dependence on the state $\rho\/$ and on
$\alpha \in SU(2)\/$ being clearly separated.
The state dependent variables $s$ and
$\tilde{u}$ transform under $SU(2)\/$ like a
scalar and a cartesian vector respectively,
and can be evaluated from the equation
\begin{equation}
Tr(\rho a^{\dagger}_r a_s)=s\delta_{rs}+u_j(\sigma_j)_{rs}
\quad r,s=1,2.
\end{equation}
The term involving the expectation values of quartics in
$a_r\/$ and $a_r^{\dagger}\/$ can be written
in terms of the non-compact generators
$\tilde{K}\/$ and $\tilde{L}\/$ of $Sp(4,\Re)\/$,
and a vector $\tilde{\lambda}\/$
representing the $SU(2)\/$ element involved:
\begin{eqnarray}
Tr(\rho a(\alpha)^{{\dagger}^{2}}a(\alpha)^2)&=&
\frac{1}{4}\lambda_j \lambda^{\star}_k H_{jk},\nonumber \\
H_{jk}=H_{kj}^{\star}&=&Tr(\rho (K_j-i
L_j)(K_k+iL_k)), \quad j,k=1,2,3\, ,\nonumber \\
\tilde{\lambda}=\tilde{\lambda}(\alpha)&=&
-i\alpha^{T}\sigma_2\tilde{\sigma}\alpha,\quad
  \tilde{\lambda}(\alpha).\tilde{\lambda}(\alpha)=0.
\end{eqnarray}
The hermitian matrix $H\/$ can be written in terms of two
real matrices, the real symmetric $R\/$ and real
antisymmetric $S\/$, as $H=R+iS\/$. The
matrix $R\/$ transforms under $SU(2)\/$ as a
second rank tensor whereas the
matrix $S\/$ can be represented by a cartesian
vector $\tilde{v}\/$ under $SU(2)\/$,
related to $S$ by $v_j=\frac{1}{2}\epsilon_{jkl}S_{kl}$.

The denominator of $Q(\rho;\alpha)\/$ is $U(2)\/$ invariant
since  the operator ${\cal A}^{\dagger}{\cal A}=
a_1^{\dagger}a_1+a_2^{\dagger}a_2\/$
is $U(2)\/$ invariant; it does not depend upon $\alpha$
and can be written in terms of $s\/$ as:
\begin{equation}
{\cal D}(Q(\rho;\alpha))=
Tr(\rho{\cal A}^{\dagger}{\cal A})=2 s
\end{equation}
After some algebra, the complex vector $\tilde{\lambda}\/$
can be eliminated in favour of the real vector $\tilde{q}\/$, and
$Q(\rho;\alpha) \/$ can be written
in terms of the state dependent symmetric second
rank tensor $R\/$, the vectors $\tilde{u},\tilde{v}\/$
and the scalar $s\/$ as:
\begin{equation}
Q(\rho;\alpha)=Q(\rho;\tilde{q}(\alpha))=
\frac{1}{8 s}(Tr R - q_j q_k R_{jk} +
2 \tilde{v}.\tilde{q} -
4 (s + \tilde{u}.\tilde{q})^{2})
\label{Q-alpha}
\end{equation}
Using the $U(2)\/$ covariance of $Q(\rho;\alpha)$,
we can assume without loss of generality that
the real symmetric matrix $R\/$ is diagonal, and
eq(\ref{Q-alpha}) then takes the simpler form:
\begin{equation}
Q(\rho;\tilde{q}(\alpha))=\frac{1}{8 s}(Tr R -
\sum_{j}q_{j}^{2}R_{jj} + 2 \tilde{v}.\tilde{q}
-4 (s + \tilde{u}.\tilde{q})^{2})
\end{equation}
The dependence of $Q(\rho;\tilde{q}(\alpha))\/$ on
$\alpha \in SU(2)\/$ is through the real unit vector
$\tilde{q}(\alpha)\/$, which can be represented on
the surface of a unit sphere. In order to obtain
the invariant Mandel parameter $Q(\rho)\/$ for a given
two-mode state, we have to minimize
$Q(\rho;\tilde{q}(\alpha))\/$ with respect to
$\tilde{q}(\alpha)\/$, the parameters
$R,\tilde{v},s,\tilde{u}\/$ being determined
by $\rho\/$. The most convenient coordinates which one chooses
on the surface of the sphere to carry out this
minimization will depend upon the physical
state $\rho\/$ under consideration.
\section{Application to Two-mode Squeezed Coherent States,
squeezed thermal states and superposition of Coherent states}
\label{sq-coh}
\setcounter{equation}{0}
In this Section, we apply the formalism developed in
the previous Section to various interesting two-mode states.
Here we will see the relation with the classification of
two-mode squeezing transformations given in~\cite{two-mode}.
\subsection{The case of squeezed coherent states}
The most general squeezed coherent state is obtained
by applying the operator ${\cal U}(\tilde{k},\tilde{l}) =
e^{i(\tilde{k}.\tilde{K}+
\tilde{l}.\tilde{L})}\/$ to the two-mode coherent
state $\vert z_1,z_2\rangle\/$ defined in
eqn.~(\ref{coh-st-def}), for some complex
$z_1,z_2\/$, where $\tilde{K}$ and $\tilde{L}\/$ are
the non-compact generators of $Sp(4,\Re)\/$ defined in
eqn.~(\ref{sp4r-gen}) and $\tilde{k}$ and
$\tilde{l}\/$ are real vectors. The operator
${\cal U}(\tilde{k},\tilde{l})\/$ is conjugate
to ${\cal U}^{(0)}(a,b)=\exp i(aK_2+bL_1)\/$ for some
$a\geq b\geq 0\/$, via an operator ${\cal U}(S^{c}(U))\/$:
\begin{eqnarray}
{\cal U}(\tilde{k},\tilde{l})&=&
{\cal U}^{-1}(S^{(c)}(U)){\cal U}^{(0)}(a,b){\cal U}(S^{(c)}(U)),
\nonumber \\
{\cal U}^{(0)}(a,b)&=&
\exp{\left(\frac{(a-b)}{4}
\left(a_1^{{\dagger}^{2}}-a_1^2\right)\right)}.
\exp{\left(\frac{(a+b)}{4}
\left(a_2^{{\dagger}^{2}}-a_2^2\right)\right)}
\label{def-uab}
\end{eqnarray}
Each ${\cal U}^{(0)}(a,b)\/$ is a representative of an
equivalence class of two-mode squeezing transformations. For
$a=b\/$ we have the essentially single mode case, while
for $b=0\/$ we have maximal involvement of the two modes.
For the minimization of $U(2)\/$ covariant
$Q(\rho;\alpha)\/$, the overall $U(2)\/$ factor
${\cal U}^{-1}(S^{(c)}(U)\/$ is irrelevant.
Also, the action of the operator
${\cal U}(S^{(c)}(U))\/$ on $\vert z_1,z_2\rangle\/$
transforms it into another coherent state
$\vert z_1^\prime ,z_2^\prime\rangle\/$,
with $z_1^\prime , z_2^\prime$ related
to $z_1,z_2\/$ through the corresponding $U(2)\/$
transformation. Thus it suffices
to examine the particular class of
squeezed coherent states
\begin{equation}
\vert z_1,z_2, a, b\rangle = {\cal U}^{(0)}(a,b)\;\vert
z_1, z_2 \rangle.
\label{sqcoh-states}
\end{equation}
A complete discussion of the two-mode squeezing
transformations and squeezed states has been
given in~\cite{two-mode}.

The Mandel parameter $Q(z_1,z_2,a,b;\tilde{q}(\alpha))\/$
for the $SU(2)\/$ transformed mode for  squeezed
coherent states can be calculated by straightforward
algebra and turns out to be rather lengthy. The
complete expression is given
in the appendix (eqn.~(\ref{Q-sqcoh})).
$Q(z_1,z_2,a,b;\tilde{q}(\alpha))\/$ depends on $a,b\/$
through hyperbolic functions and on $\vert z_1
\vert, \vert z_2 \vert$ through polynomial functions.
Its dependence on the phases of $z_1\/$ and $z_2\/$
and the polar coordinates on the surface of the unit
sphere describing the unit vector $\tilde{q}(\alpha)$,
is through trigonometric functions and is
oscillatory in nature.
In order to obtain the invariant Mandel parameter, this function
has to be  minimized with respect to
$\tilde{q}(\alpha)\/$. Since this is not possible
analytically, the results obtained numerically
are displayed in Figures~(\ref{sq-coh-1}), (\ref{sq-coh-2}) and
(\ref{sq-coh-3})~\cite{scale}.

In each figure, we plot the minimum value of
$Q(z_1,z_2,a,b;\tilde{q}(\alpha))\/$ as a function of the
squeeze factors $a\/$ and $b\/$, keeping the complex
displacements $z_1\/$ and $z_2\/$ fixed.
Figure~(\ref{sq-coh-1}(a)) displays
the results for the squeezed vacuum; this never shows
SPS. The plots of Figures~(\ref{sq-coh-1}(b),(c) and
(d)) on the other hand are obtained by varying the phase of one
of the displacements ($z_2$), keeping its magnitude fixed, with the
other displacement ($z_1$) being zero. Different values for the phase of
the non-zero displacement give qualitatively different
results; in particular when this phase
is $\frac{\pi}{2}\/$, as is clear from
Figure~(\ref{sq-coh-1}(d)) even some of the essentially
single mode states lying along $a=b\/$ show SPS. In
Figure~(\ref{sq-coh-2}) we choose equal magnitudes of
displacements for the two modes; plots have been generated
for different values of their phases.
The displacement parameters in Figure~(\ref{sq-coh-3}) are
unequal in magnitude; four plots have been given for
the same choices of phase values as in the corresponding
plots in Figure~(\ref{sq-coh-2}). The qualitative
features of individual plots are similar to the
corresponding plots in Figure~(\ref{sq-coh-2})
though the actual values of the invariant Mandel
parameter are different.

We now make some general remarks about the results described
above. In all the plots of the three Figures~(\ref{sq-coh-1}),
(\ref{sq-coh-2}) and (\ref{sq-coh-3}), every point in
the region $b > a\/$ can be mapped onto a corresponding
unique point in some region $a > b\/$
(which is in general not in the same figure),
through that $U(2)\/$ transformation of the displacements
$z_1\/$ and $z_2\/$, which effectively changes
${\cal U}^{(0)}(a,b)\/$ to ${\cal U}^{(0)}(b,a)$.
Whenever the displacement parameters are invariant under
this particular $U(2)\/$ transformation, the plot has a
symmetry about the line $a=b\/$;
as in all the plots of Figure~(\ref{sq-coh-1}). Such a symmetry
is not exhibited by the plots of Figures~(\ref{sq-coh-2})
and (\ref{sq-coh-3}).
In all the plots the invariant Mandel parameter
is zero or negative along the line $a=b\/$ i.e.
for the subset of essentially single
mode squeezed states. This happens because, even though the
choice of displacement parameters is such that
the single mode which is squeezed has superpoissonian statistics,
$(Q>0)$, the minimisation chooses the other mode which
is in a coherent state $(Q=0)$. Apart from the case of
squeezed vacuum(Figure~(\ref{sq-coh-1}a)) all
other choices of displacement show SPS
for some values of the squeeze parameters $a,b\/$.
When squeezing becomes large in comparison to the
displacement, and we are away from the line $a=b\/$,
SPS disappears and the states tend to become
more and more superpoissonian.
\subsection{The case of squeezed thermal states}
We next look at the case of a two-mode isotropic
thermal state subjected to squeezing. The normalized
density operator corresponding to the inverse
temperature $\beta = \hbar\omega/kT\/$ is explicitly
$U(2)\/$ invariant and described in
the Fock representation by:
\begin{eqnarray}
\rho_0(\beta) &=& (1 - e^{-\beta})^2\;
\exp\left[-\beta(a_1^\dagger\:
a_1 + a_2^\dagger\:a_2)\right]\nonumber\\ & = & (1 -
e^{-\beta})^2\;\sum_{n_1,n_2=0}^{\infty}
e^{-\beta (n_1 + n_2)}
\vert n_1,n_2\rangle\langle n_1,n_2\vert,
\end{eqnarray}
with $U(2)\/$ invariance expressed by:
\begin{eqnarray}
e^{\textstyle i\theta J_{0}}\;\rho_0(\beta)\;
e^{\textstyle -i\theta J_{0}}&
=& e^{\textstyle i\vec{\alpha}\cdot\vec{J}} \;\rho_0(\beta)\;
e^{\textstyle -i\vec{\alpha}\cdot\vec{J}} = \rho_0(\beta).
\label{rhoth-invar}
\end{eqnarray}
Therefore it suffices to examine the properties of the density
operator obtained by conjugating $\rho_0(\beta)\/$ with ${\cal
U}^{(0)}(a,b)\/$:
\begin{equation}
\rho(\beta;a,b) = {\cal U}^{(0)}(a,b)\;\rho_0(\beta)\;
{\cal U}^{(0)}(a,b)^{-1}.
\label{rho-betaab}
\end{equation}

The Mandel parameter $Q(\beta;a,b;\tilde{q}(\alpha))\/$
for the state $\rho(\beta;a,b)\/$ is calculable by
straightforward algebra:
\begin{eqnarray}
&&Q(\beta;a,b;\tilde{q}(\alpha))=\nonumber \\
&&\left[(e^{\beta} -1)\,(2\, (1 - e^{\beta}) + 2\, (1 + e^{\beta})
\,\cosh(2 a) \,\cosh(2b))\right]^{-1} \times
\nonumber \\
&&\left[ \frac{1}{4}\left((1 - q_3^2)\,
(2\,(1 - e^{\beta})^2 + 4\, (1 - e^{2 {\beta}}) \,
\cosh(2 a) \,\cosh(2 b) +
(1 + e^{\beta})^2\, (\,\cosh(4 a) + \,\cosh(4 b)))  \right.\right.
\nonumber \\ &&
\,+\,(1 + e^{\beta})^2\, ({q_1}^2 - {q_2}^2)
(\,\cosh(4 a) - \,\cosh(4 b))
\nonumber \\&& \left.
    - \frac{1}{2}\,(1 + {q_3}^2)\,
(10 - 12 e^{\beta} + 10 e^{2 {\beta}} +
            16\, (1 - e^{2 {\beta}}) \,\cosh(2 a) \,\cosh(2 b) +
            6\, (1 + e^{\beta})^2 \,\cosh(4 a) \,\cosh(4 b))\right)
\nonumber \\ &&
- \frac{1}{2}((1 + e^{\beta})\, q_3\,
(4 - 4 e^{\beta} + 6 (1 + e^{\beta}) \,\cosh(2 a)
\,\cosh(2 b)) \,{\rm sinh}(2 a)
     \,{\rm sinh}(2 b))
\nonumber \\ &&
\left.
- \frac{1}{2}((2 - 2 e^{\beta} +
2 (-1 + e^{2{\beta}}) \,\cosh(2 a) \,\cosh(2 b)) +
     ((1 + e^{\beta}) \,q_3 \,
{\rm sinh}(2 a) \,{\rm sinh}(2 b)))^2 \right] \nonumber \\
\end{eqnarray}
here $q_1,q_2,q_3\/$ are the cartesian components of
$\tilde{q}\/$ with $q_1^2+q_2^2+q_3^2=1\/$.

The minimum value of the function
$Q(\beta,a,b,\tilde{q}(\alpha))\/$,
the parameter $Q(\rho(\beta,a,b))\/$,
can be calculated analytically. The state $\rho(\beta,a,b)\/$
being the squeezed thermal state is always superpoissonian.
For a given temperature(given $\beta$) this
superpoissonian nature is least for the case when only
one mode is squeezed($a=b$),
increases as the squeezing becomes
increasingly two mode in nature and
finally is maximum when the
state is maximally two mode squeezed i.e.
when $a=0\,(b=0)\/$ for a given $b\,(a)\/$.
When the temperature is changed
the states with higher temperatures (lower $\beta$)
are more superpoissonian compared to the ones at lower
temperatures (higher $\beta$).Thus for fixed $a\/$ and $b\/$,
$Q(\beta;a,b)$ increases as $\beta\/$ decreases. The actual plots
of $Q(\rho(\beta,a,b))\/$ as a function of $a,b$ are given at
different temperatures in
Figure(~\ref{sq-thermal-plot})~\cite{scale}.

It is interesting to note that the particular mode
for which the function $Q(\rho(\beta,a,b),
\tilde{q}(\alpha))\/$ is minimum turns out to be one of
the original modes, corresponding to $q_3=\pm 1$.
This happens because the thermal state density matrix
$\rho_0(\beta)\/$ is explicitly $U(2)\/$
invariant and the representative two-mode squeezing
operator ${\cal U}^{(0)}(a,b)\/$ can be factorized into two
commuting operators, each pertaining to one of the
original modes see eq.~(\ref{def-uab}).
In general, for a different choice of the
representative operators, the minima could occur at an
arbitrary $SU(2)\/$ transformed first mode.
All the plots of Figure(~\ref{sq-thermal-plot}) are symmetric
about the line $a=b\/$ because of the explicit $U(2)\/$
invariance of the thermal state density
matrix $\rho_0(\beta)\/$ (eqn.(~\ref{rhoth-invar})).
\subsection{The case of superposition of coherent states}
Lastly we apply our formalism to the superposition of two
two-mode coherent states. In this case, no squeezing
transformation ${\cal U}^{(0)}(a,b)\/$ is involved.
For simplicity we consider only the case with
real displacements.

A general superposition of two two-mode coherent states with
real displacements and a phase difference $\eta\/$ between
them is given by:
\begin{eqnarray}
\label{def-sup-coh}
\vert \psi(u_1,u_2,v_1,v_2,r,\eta)\rangle
&=& \frac{1}{N}(\vert u_1,u_2\rangle + r
\exp{(i\eta)} \vert v_1,v_2 \rangle) \nonumber \\
\mbox{where}\quad N^2 &=&1+r^2+2\,r \,\cos{\eta}
\exp{\left(-\frac{1}{2}(u_1^2 + u_2^2 + v_1^2 + v_2^2)+
u_1\,v_1+u_2\,v_2 \right)}
\label{sup-coh-def}
\end{eqnarray}
With the help of a $U(2)\/$ transformation, without any loss of
generality we can set $v_2=0\/$ and thus it suffices to study
only the states $\vert \psi(u_1,u_2,v_1,0,r,\eta)\rangle\/$.
The Mandel parameter for the $SU(2)\/$ transformed mode,
the function $Q(u_1,u_2,v_1,r,\eta;\tilde{q}(\alpha))\/$
for this superposition of coherent states is given in
terms of the polar coordinates $\theta\/$ and
$\phi\/$ on the surface of the sphere representing
$\tilde{q}\/$ as:
\begin{eqnarray}
&& Q(u_1,u_2,v_1,r,\eta;\tilde{q}(\alpha))=
\nonumber \\&&
\left[4\left( u_1^2 + u_2^2 + {r^2}\, v_1^2 +
  2\,{e{}^{-\frac{1}{2}({ u_1}^2 + { u_2}^2
 +{ v_1}^2) + { u_1}\,{ v_1}}}\,r\,{ u_1}\,
 { v_1}\,\cos ({ \eta})\right) \right]^{-1} \times
\nonumber \\
&& \left[ \left( 1 + {r^2} + 2\,{e^
{-\frac{1}{2}({{{ u_2}}^2}+{\left( { u_1} - { v_1} \right)
}^2)}}\,r\,
    \cos ({ \eta}) \right) \times  \right.
\nonumber \\&&
   \left( 4\,\left( {{{ u_1}}^4} + {r^2}\,{{{ v_1}}^4} \right) \,
   {{\cos ({\theta\over 2})}^4} +
   4\,{{{ u_2}}^4}\,{{\sin ({\theta\over 2})}^4}+
   8\,{{{ u_1}}^3}\,{ u_2}\,
   {{\cos ({\theta\over 2})}^2}\,\cos (\phi)\,
  \sin (\theta) \right.
\nonumber \\ && \left.\quad \quad \quad +
    8\,{ u_1}\,{{{ u_2}}^3}\,\cos (\phi)\,
    {{\sin ({\theta\over 2})}^2}\,\sin (\theta) +
    2\,{{{ u_1}}^2}\,{{{ u_2}}^2}\,
    \left( 2 + \cos (2\,\phi) \right) \,
         {{\sin (\theta)}^2} \right)
\nonumber \\ && +
  r\,\left( e^{-\frac{1}{2}({ u_2}^2 +
       \left( {u_1} - {v_1} \right)^2)}\,
       \left( 1 + {r^2} \right)  +
       2\,{e^{-{{{ u_2}}^2} -
      {{\left( { u_1} - { v_1} \right) }^2}}}\,
         r\,\cos ({ \eta}) \right) \times
\nonumber \\ &&
   \left( 8\,{{{ u_1}}^2}\,{{{ v_1}}^2}\,
         {{\cos ({\theta\over 2})}^4}\,
         \cos ({ \eta}) \right.
\nonumber \\ &&\left.\quad\quad +
       8\,{ u_1}\,{ u_2}\,{{{ v_1}}^2}\,
         {{\cos ({\theta\over 2})}^2}\,\cos
        ({ \eta} + \phi)\,\sin (\theta) +
        2\,{{{ u_2}}^2}\,{{{ v_1}}^2}\,\cos ({ \eta} + 2\,\phi)\,
         {{\sin (\theta)}^2} \right)
\nonumber \\ && -
  \left( 2\,\left( {{{ u_1}}^2} + {r^2}\,{{{ v_1}}^2} \right) \,
            {{\cos ({\theta\over 2})}^2} +
           2\,{{{ u_2}}^2}\,{{\sin ({\theta\over 2})}^2} +
           2\,{ u_1}\,{ u_2}\,\cos (\phi)\,\sin (\theta)  \right.
\nonumber \\ &&
     \left. \left. \quad +\frac{1}{2} {e^{-\frac{1}{2}({{u_2}^2} +
        2\,{{\left( { u_1} - { v_1} \right) }^2})}}\,r\,
      \left( 4\,{ u_1}\,{ v_1}\,{{\cos ({\theta\over 2})}^2}\,
      \cos ({ \eta}) + 2\,{ u_2}\,{ v_1}\,
\cos ({ \eta} + \phi)\,\sin (\theta) \right) \right)^2\right]
\end{eqnarray}
The minimum values of this function with respect to $\theta\/$
and $\phi\/$ have been computed numerically and the results are
shown in Figure~(\ref{sup-coh-plot}). Each plot in this figure
contains two curves showing $Q(\rho)\/$ as a function of the
relative phase $\eta\/$ corresponding to two different values of
relative weight factor $r\/$. The amount of SPS varies with
the relative phase in a similar way for all the plots. For all
parameter values in all plots $Q(\rho)\leq0\/$. This happens
because the most general superposition of two two-mode coherent
states can be transformed with the help of a $U(2)\/$
transformation into a product state with one factor
being a coherent state, and the other a
superposition of two one-mode coherent states:
\begin{equation}
 \frac{1}{N}(\vert u_1^{\prime} \rangle + r
\exp{(i\eta)} \vert v_1^{\prime}\rangle)\vert v_2^{\prime}
\rangle = \frac{1}{N}{\cal U}(S^{(c)}(U))(\vert u_1,u_2\rangle + r
\exp{(i\eta)} \vert v_1,v_2 \rangle)
\label{sup-product}
\end{equation}
Thus when $Q(\rho;\alpha)\/$ turns out to be nowhere negative
the minimization chooses that $U(2)\/$ transformed  mode which
is in a coherent state.

It is interesting to point out that for a factorized two-mode
state such as the expression on the left hand side of
eqn.~(\ref{sup-product}), the mode $a(\overline{\alpha})\/$which
minimizes $Q(\rho;\alpha)\/$ is generally neither of the two
initial modes but a nontrivial combination of them.
\section{Concluding Remarks}
\label{con-remarks}
\setcounter{equation}{0}
The main aim of this paper has been to develop a specific
signature of non-classicality for two-mode states. Both
quadrature squeezing and SPS are well defined concepts
for a single mode. In this paper we have extended the notion of
SPS to two modes by showing how to choose the
appropriate single mode which shows SPS to the maximum
extent, considering all modes related to each other by passive
$U(2)\/$ transformations as equivalent. A similar treatment of
quadrature squeezing has been given elsewhere.

We would like to emphasize the subtle role played by the
choice of the denominator of $Q(\rho;\alpha)\/$.
Any choice which is everywhere non-negative will
not change the qualitative results
obtained from the minimization of $Q(\rho;\alpha)\/$
i.e. the super or subpoissonian nature of the state $\rho$.
However the extent of SPS, and the location
of the most non-classical mode, depend upon the exact
choice one makes for the
denominator. To illustrate this point we choose
the two-mode Fock state $\vert n_1,n_2\rangle\/$.
The Mandel parameter for the $SU(2)\/$
transformed mode is given by:
\begin{eqnarray}
Q(n_1,n_2,\tilde{q}(\alpha))=
&&\frac{1}{4\,\left(n_1 +  n_2 \right) }\times
\left( -2\,(n_1 + n_2) + (n_1 + n_2)^2 +
\right.
\nonumber \\
&&\!\!\!\! \left.
 \left(n_1(1 - n_1) + n_2(1 - n_2)\,\right)
   \left( {q_1}^2 + {q_2}^2 \right)
 - 2\,\left(n_1 - n_2 \right)\,{q_3} -
\left( n_1 + n_2 \right)^2\,{q_3}^2 \right)
\end{eqnarray}
This function reaches its minimum at $q_3=+1\/$ with
minimum value $-\frac{\textstyle{n_1}}{\textstyle{n_1+n_2}}\/$
for $n_1 > n_2\/$ and at $q_3=-1\/$ with the minimum value
$-\frac{\textstyle{n_2}}{\textstyle{n_1+n_2}}\/$ for
$n_2 > n_1\/$. Thus for our $U(2)\/$ invariant choice of
the denominator $Tr(\rho {\cal A}^{\dagger}{\cal A})\/$,
for a Fock state, the mode with the larger number of photons
is more non-classical. On the other hand if one chooses the
$U(2)\/$ covariant denominator
$Tr(\rho a(\alpha)^{\dagger}a(\alpha))\/$, for a Fock state,
both the modes are equally non-classical irrespective of the
number of photons present in each mode: the minimum value of
this alternatively defined parameter is $-1\/$ for each mode.

 For one-mode  fields the Mandel parameter can be
written as a function of the number operator
$a^{\dagger}a\/$ and hence is determined by
(the moments of) the photon number distribution.
In contrast, for two-mode fields the  Mandel
parameter for the $SU(2)\/$ transformed mode can not
be expressed as a function of the number
operators $a_1^{\dagger}a_1$ and $a_2^{\dagger}a_2\/$ and
therefore is not determined by the photon number distributions
in the original modes. There could be other signatures
of non-classicality which are meaningful at the one-mode level
and can be extended in the spirit of this paper to
more than one mode. In contrast, it will be interesting
to explore the possibility of having
signatures of non-classicality which are not definable at the
one-mode level at all, but are present only at the two-mode
level. These will be presented elsewhere.
\newpage
\appendix
\section{}
We give here the function $Q(z_1,z_2,a,b;\alpha)\/$ for the
squeezed coherent state with $z_1=u e^{i \varphi_u}$ and
$z_2= v e^{i \varphi_v}$. The first term is the denominator,
followed by the numerator terms arranged according
to their dependence on $a\/$ and $b\/$. First the
terms independent of $a,b\/$ appear, followed by
the ones depending upon $a\/$ or $b\/$ alone, and
then the ones depending on both $a\/$ and $b$.
The last three terms originate from quadratic
expressions of creation and annihilation
operators and are not arranged.
\begin{eqnarray}
\label{Q-sqcoh}
&&Q(\alpha;z_1,z_2,a,b)= \nonumber \\
 && 2 \left[-2 + \cosh (2\,\left( a - b \right) ) +
     2\,{u^2}\,\cosh (2\,\left( a - b \right) ) +
     \cosh (2\,\left( a + b \right) ) +
     2\,{v^2}\,\cosh (2\,\left( a + b \right) ) \right.
\nonumber \\&&   +
\left. 2\,{u^2}\,\cos (2\,{ \varphi_u})\,
{\rm Sinh}(2\,\left( a - b \right) ) +
2\,{v^2}\,\cos (2\,{ \varphi_v})\,{\rm Sinh}(2\,\left( a +
b\right) )\right]^{-1}\times
\nonumber \\ && \left[ \begin{array}{c}\mbox{}\\
\mbox{}\end{array} \right.
  \frac{1}{8}\left(5 + {u^4} + {v^4} + 2\,
\left( {u^2} + {v^2} \right) \right)
\nonumber \\ && +
\frac{1}{8}\left({v^4}\,
\cos (4\,{ \varphi_v})\,\left( -1 + \cos (\theta ) \right)
\right)
\nonumber \\ && +
\frac{1}{8}\left(\left( {u^2} - {v^2} \right) \,
\left( 2 + {u^2} + {v^2} \right) \,
\cos (\theta )\right)
\nonumber \\ && -
\frac{1}{8}\left({u^4}\,
\cos (4\,{ \varphi_u})\,
\left( 1 + \cos (\theta ) \right) \right)
 \nonumber \\ && -
 \frac{1}{16} \left(
\left( 1 + {u^2}\,\left( 2 +
{u^2} - {u^2}\,\cos (4\,{ \varphi_u}) \right)  +
{v^2}\,\left( 2 + {v^2} - {v^2}\,
\cos (4\,{ \varphi_v}) \right)
\right) \,{{\sin (\theta )}^2} \right)
\nonumber \\ && +
\frac{1}{4}\left(u\,v\,\cosh (2a)\,
\left(\begin{array}{c} -8\,\cos (\phi )\,
 \cos ({ \varphi_u} - { \varphi_v}) -
   \left( 2 + {u^2} + {v^2} +
  \left( {u^2} - {v^2} \right) \,\cos (\theta ) \right) \,
  \sin (\phi )\,\sin ({ \varphi_u} - { \varphi_v}) \\
	- {u^2}\,\left( 1 + \cos (\theta ) \right) \,
	 \sin (\phi )\,
	 \sin (3\,{ \varphi_u} + { \varphi_v}) +
	{v^2}\,\left( 1 - \cos (\theta ) \right) \,\sin (\phi )\,
\sin ({ \varphi_u} + 3\,{ \varphi_v})\end{array} \right) \right) \,
\sin (\theta )
\nonumber \\ && +
\frac{1}{4}\left( u\,v\,
\left(\begin{array}{c} -8\,\cos (\phi)\,
\cos ({ \varphi_u} + { \varphi_v}) -
\left( {u^2} - {v^2} + \left( 2 + {u^2} + {v^2} \right)\,
\cos (\theta ) \right) \,\sin (\phi )\,
\sin ({ \varphi_u} + { \varphi_v}) \\ -
{v^2}\,\left( 1 - \cos (\theta ) \right) \,\sin (\phi )\,
   \sin ({ \varphi_u} - 3\,{ \varphi_v}) -
 {u^2}\,\left( 1 + \cos (\theta ) \right) \,\sin (\phi )\,
	 \sin (3\,{ \varphi_u} - { \varphi_v})
\end{array} \right) \,\sin (\theta )\,{\rm Sinh}(2a) \right)
\nonumber \\ && +
\frac{1}{4}\left( u\,v\,\cosh (-2b)\,
\left( \begin{array}{c} 8\,\sin (\phi )\,
\sin ({ \varphi_u} - { \varphi_v})\,
\sin (\theta )+ \cos (\phi ) \,
\cos ({ \varphi_u} - { \varphi_v})\,\left( 2 + {u^2} + {v^2} +
\left( {u^2} - {v^2} \right) \,
\cos (\theta ) \right) \\- \cos (\phi )\,
2\,{u^2}\,\cos (3\,{ \varphi_u} + { \varphi_v})\,
{{\cos ({{\theta }\over 2})}^2}
- 2\,{v^2}\,\cos (\phi )\, \cos ({ \varphi_u} + 3\,{ \varphi_v})\,
 {{\sin ({{\theta }\over 2})}^2}
 \,\sin (\theta ) \end{array}\right)
\right)
\nonumber \\ && +
\frac{1}{4}
\left( u\,v\,\left(\begin{array}{c} {v^2}\,
\cos (\phi )\,\cos ({ \varphi_u} - 3\,{ \varphi_v})\,
\left( -1 + \cos (\theta ) \right)  +
{u^2}\,\cos (\phi )\,\cos (3\,{ \varphi_u} - { \varphi_v})\,
\left( 1 + \cos (\theta ) \right)  \\ -
 \cos (\phi )\,\cos ({ \varphi_u} + { \varphi_v})\,
 \left( {u^2} - {v^2} +
\left( 2 + {u^2} + {v^2} \right) \,
   \cos (\theta ) \right)  -
  8\,\sin (\phi )\,\sin ({ \varphi_u} + { \varphi_v})
\end{array} \right) \,\sin (\theta )\,
     {\rm Sinh}(-2b) \right)
\nonumber \\ && +   \frac{1}{8} \left( 3 + 12\,{u^2} + 6\,{u^4} +
	2\,{u^4}\,\cos (4\,{ \varphi_u}) \right) \,
     {{\cos ({{\theta }\over 2})}^4}\,\cosh (4 (a\,-\,b))
\nonumber \\ && +
\frac{1}{8}\left( 3 + 12\,{v^2} + 6\,{v^4} +
	2\,{v^4}\,\cos (4\,{ \varphi_v}) \right) \,
     \cosh (4 (a\,+\,b))\,{{\sin ({{\theta }\over 2})}^4}
\nonumber \\ && +
\frac{1}{2}\left({u^2}\,\left( 3 +
     2\,{u^2} \right) \,\cos (2\,{ \varphi_u})\,
     {{\cos ({{\theta }\over 2})}^4}\,{\rm Sinh}(4 (a\,-\,b)) \right)
\nonumber \\ && +
\frac{1}{2}\left({v^2}\,\left( 3 + 2\,{v^2} \right)
  \,\cos (2\,{ \varphi_v})\,{{\sin ({{\theta }\over 2})}^4}\,
   {\rm Sinh}(4 (a\,+\,b)) \right)
\nonumber \\ && +
\cosh (2(a - b))\,
\left( -\left( \left( 1 + 2\,{u^2} \right) \,
{{\cos ({{\theta }\over 2})}^2} \right)  +
{{{u^2}\,{v^2}\,\cos (2\,{ \varphi_u})\,
\sin (2\,\phi )\,\sin (2\,{ \varphi_v})\,
{{\sin (\theta )}^2}}\over 2} \right)
\nonumber \\ && +
\cosh (2(a - b))\,\left(
 - \left( \left( 1 + 2\,{v^2} \right) \,
 {{\sin ({{\theta }\over 2})}^2} \right)  -
 {{{u^2}\,{v^2}\,\cos (2\,{ \varphi_v})\,
\sin (2\,\phi )\,\sin (2\,{ \varphi_u})\,
 {{\sin (\theta )}^2}}\over 2} \right)
\nonumber \\ && +
\left( -\left( {u^2}\,\cos (2\,{ \varphi_u})\,
\left( 1 + \cos (\theta ) \right)  \right)  +
{{\left( 1 + 2\,{u^2} \right) \,{v^2}\,\sin (2\,\phi )\,
\sin (2\,{ \varphi_v})\,
{{\sin (\theta )}^2}}\over 4} \right) \,{\rm Sinh}(2(a - b))
\nonumber \\ && +
\left( {v^2}\,\cos (2\,{ \varphi_v})\,
\left( -1 + \cos (\theta ) \right)  -
{{{u^2}\,\left( 1 + 2\,{v^2} \right) \,\sin (2\,\phi )\,
\sin (2\,{ \varphi_u})\,{{\sin (\theta )}^2}}\over 4}
\right) \,{\rm Sinh}(2(a - b))
\nonumber \\ && +
\frac{1}{4}\left(
\left( \left( 1 + 2\,{u^2} \right) \,
\left( 1 + 2\,{v^2} \right)  +
2\,{u^2}\,{v^2}\,\cos (2\,\phi )\,\cos (2\,{ \varphi_u})\,
\cos (2\,{ \varphi_v}) \right) \,
\cosh (2(a - b))\,\cosh (2(a - b))\,
{{\sin (\theta )}^2} \right)
\nonumber \\ && +
\frac{1}{8}\left(\left( 4\,{u^2}\,
\left( 1 + 2\,{v^2} \right) \,\cos (2\,{ \varphi_u}) +
2\,\left( 1 + 2\,{u^2} \right) \,{v^2}\,\cos (2\,\phi )\,
\cos (2\,{ \varphi_v}) \right) \,
\cosh (2(a - b))\,{{\sin (\theta )}^2}\,
{\rm Sinh}(2(a - b)) \right)
\nonumber \\ && +
\frac{1}{8}\left(\left(
\left( 1 + 2\,{u^2} \right) \,
\left( 1 + 2\,{v^2} \right) \,
\cos (2\,\phi ) + 8\,{u^2}\,{v^2}\,\cos (2\,{ \varphi_u})\,
\cos (2\,{ \varphi_v}) \right) \,
{{\sin (\theta )}^2}\,{\rm Sinh}(2(a - b))\,
{\rm Sinh}(2(a - b)) \right)
\nonumber \\ && +
\frac{1}{8}\left(\left( 2\,{u^2}\,
\left( 1 + 2\,{v^2} \right) \,\cos (2\,\phi )\,
 \cos (2\,{ \varphi_u}) + 4\,\left( 1 + 2\,{u^2} \right) \,{v^2}\,
 \cos (2\,{ \varphi_v}) \right) \,\cosh (2(a - b))\,
{{\sin (\theta )}^2}\,
{\rm Sinh}(2(a - b)) \right)
\nonumber \\ && +
\frac{1}{2} \left( u\,v\,\cos (\phi )\,
    \left( 3\,\left( 1 + {u^2} \right) \,
    \cos ({ \varphi_u} - { \varphi_v}) + {u^2}\,
     \cos (3\,{ \varphi_u} + { \varphi_v}) \right) \,
     {{\cos ({{\theta }\over 2})}^2}\,
   \cosh (2(2 a - b))\,\sin (\theta )
     \right)
\nonumber \\ && +
\frac{1}{2} \left(u\,v\,\cos (\phi )\,
   \left( {u^2}\,\cos (3\,{ \varphi_u} - { \varphi_v}) +
	3\,\left( 1 + {u^2} \right) \,\cos ({ \varphi_u} +
{ \varphi_v}) \right) \,
{{\cos ({{\theta }\over 2})}^2}\,\sin (\theta )\,{\rm
Sinh}(2(2 a - b)) \right)
\nonumber \\ && +
\frac{1}{2} \left(u\,v\,
\cos (\phi )\,\left( 3\,\left( 1 + {v^2} \right) \,
    \cos ({ \varphi_u} - { \varphi_v}) + {v^2}\,
\cos ({ \varphi_u} + 3\,{ \varphi_v}) \right)
\,\cosh (2(2 a + b))\,
{{\sin ({{\theta }\over 2})}^2}\,\sin (\theta )
     \right)
\nonumber \\ && +
\frac{1}{2} \left(u\,v\,\cos (\phi )\,
    \left( {v^2}\,\cos ({ \varphi_u} - 3\,{ \varphi_v}) +
	3\,\left( 1 + {v^2} \right) \,
\cos ({ \varphi_u} + { \varphi_v}) \right) \,
{{\sin ({{\theta }\over 2})}^2}\,\sin (\theta )\,{\rm
Sinh}(2(2 a + b)) \right)
\nonumber \\ && +
\frac{1}{2} \left( u\,v\,
{{\cos ({{\theta }\over 2})}^2}\,\sin (\phi )\,
     \left( -\left( {u^2}\,\sin
(3\,{ \varphi_u} - { \varphi_v}) \right)  +
	3\,\left( 1 + {u^2} \right) \,
\sin ({ \varphi_u} + { \varphi_v}) \right) \,
     \sin (\theta )\,{\rm Sinh}(2(a - 2\,b))\right)
\nonumber \\ && +
\frac{1}{2} \left( u\,v\,\sin (\phi )\,
\left( {v^2}\,\sin ({ \varphi_u} - 3\,{ \varphi_v}) +
3\,\left( 1 + {v^2} \right)
\,\sin ({ \varphi_u} + { \varphi_v}) \right)\,
{{\sin ({{\theta }\over 2})}^2}\,
\sin (\theta )\,{\rm Sinh}(-2(a + 2b))
\right)
\nonumber \\ && -
\frac{1}{2} \left( u\,v\,\cosh (-2(a + 2b))\,\sin (\phi )\,
   \left( 3\,\left( 1 + {v^2} \right) \,
\sin ({ \varphi_u} - { \varphi_v}) +
	  {v^2}\,\sin ({ \varphi_u} + 3\,{ \varphi_v}) \right) \,
	{{\sin ({{\theta }\over 2})}^2}\,
\sin (\theta ) \right)
\nonumber \\ && +
\frac{1}{2} \left(u\,v\,
{{\cos ({{\theta }\over 2})}^2}\,\cosh (2(a - 2\,b))\,\sin (\phi )\,
     \left( -3\,\left( 1 + {u^2} \right) \,
\sin ({ \varphi_u} - { \varphi_v}) +
{u^2}\,\sin (3\,{ \varphi_u} + { \varphi_v}) \right) \,
\sin (\theta ) \right)
\nonumber \\ && +
 \frac{1}{2}\left({u^2}\,{v^2}\,
\cos (2\,\phi )\,\sin (2\,{ \varphi_u})\,
\sin (2\,{ \varphi_v})\,
{{\sin (\theta )}^2} \right)
\nonumber \\ && -
\left[-{1\over 2} + {{\left( 1 + \cos (\theta ) \right) \,
\left( \left( 1 + 2\,{u^2} \right) \,\cosh (2(a - b)) +
2\,{u^2}\,\cos (2\,{ \varphi_u})\,{\rm Sinh}(2(a - b))
\right) }\over 4}
 \right]
\nonumber \\ && +
{{\left( 1 - \cos (\theta ) \right) \,
 \left( \left( 1 + 2\,{v^2} \right) \,\cosh (2(a - b)) +
 2\,{v^2}\,\cos (2\,{ \varphi_v})\,
{\rm Sinh}(2(a - b)) \right) }\over 4}
\nonumber \\ && +
  u\,v\,\sin (\theta )\,
 \left.\left. \left( \begin{array}{c}\sin (\phi )\,
\left( \cosh (-2b)\,
 \sin ({ \varphi_u} - { \varphi_v}) -
\sin ({ \varphi_u} + { \varphi_v})\,
{\rm Sinh}(-2b) \right)
\\+ \cos (\phi )\,
\left( \cos ({ \varphi_u} - { \varphi_v})\,\cosh (2a) +
\cos ({ \varphi_u} + { \varphi_v})\,
{\rm Sinh}(2a) \right) \end{array}\right) \right]^2 \right]
\nonumber
\end{eqnarray}
\begin{equation}
\end{equation}
{\large \bf Acknowledgments}\\
Arvind thanks University Grants Commission India for financial
support.

\setcounter{figure}{0}
\begin{figure}
\caption{Plots of the invariant Mandel parameter
$Q(\rho)\/$ for squeezed coherent states as a
function of squeeze parameters $a\/$ and $b\/$.
Fig.1(a) shows the plot for squeezed vacuum i.e. $z_1=z_2=0\/$.
Figures(1(b), (c) and (d)) show the plots for
$\vert z_1\vert =0,\,\vert z_2\vert =3.0\/$ and the
phase of $z_2$ taking the values $0,\, \pi/4\/$ and $\pi/2\/$
respectively.}
\label{sq-coh-1}
\end{figure}
\begin{figure}
\caption{Plots of the invariant Mandel  parameter $Q(\rho)\/$ for
squeezed coherent states as a function of squeeze
parameters $a\/$ and $b\/$ for the case when the magnitudes
of the displacements in the two modes are equal:
$\vert z_1\vert=\vert z_2\vert =2.0\/$.
The values of the phases of $z_1\/$ and
$z_2\/$ in Figures(2(a),(b),(c) and (d)) are
$(0,0),\,(0,\pi/4),\,(0,\pi/2)$ and $(\pi/2,\pi/2)$ respectively.}
 \label{sq-coh-2}
\end{figure}
\begin{figure}
\caption{Plots of the invariant Mandel parameter $Q(\rho)\/$ for
 squeezed coherent states as a function of squeeze
 parameters $a\/$ and $b\/$ for the case when the magnitudes
of the displacements in the two modes are unequal:
$\vert z_1\vert =2.0\/$ and $\vert z_2\vert =4.0\/$.
The values of the phases of
$z_1\/$ and $z_2\/$ in Figures(3(a),(b),(c) and (d)) are
$(0,0),\,(0,\pi/4),\,(0,\pi/2)$ and $(\pi/2,\pi/2)$ respectively.}
\label{sq-coh-3}
\end{figure}
\begin{figure}
\caption{Plots of the invariant Mandel parameter $Q(\rho)\/$ for
squeezed thermal states as a function of squeeze parameters
$a\/$ and $b\/$ at different inverse temperatures; $\beta\/$
takes the values $0.5, 1.0, 2.0\/$ and $4.0\/$ in
Figures(4(a),(b),(c) and (d)) respectively.}
\label{sq-thermal-plot}
\end{figure}
\begin{figure}
\caption{Plots of invariant Mandel parameter $Q(\rho)\/$ for
superposition of two two-mode coherent states as
a function of the phase difference between
the two states are shown for two different values of relative
weight, $r=0.5\/$ and $r=1.0\/$,and a given set of
displacements. Values of displacements $u_1\/$, $u_2\/$ and
$v_2\/$ are $(0.5,0.5,1.0),\quad(0.5,1.0,1.0),
\quad(1.5,1.0,1.0\/)$ and $(1.5,1.0,0.5)$ for
Figures(5(a),(b),(c), and (d)) respectively.}
\label{sup-coh-plot}
\end{figure}

\begin{references}
\bibitem{scale} Note that the subfigures in each figure
are not drawn to the same scale.
\end{references}
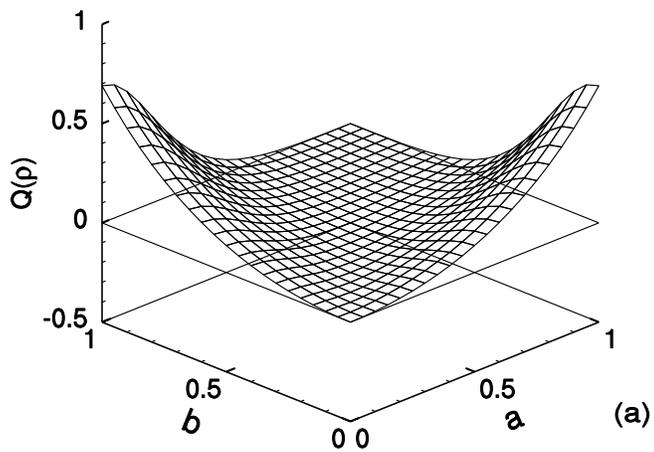
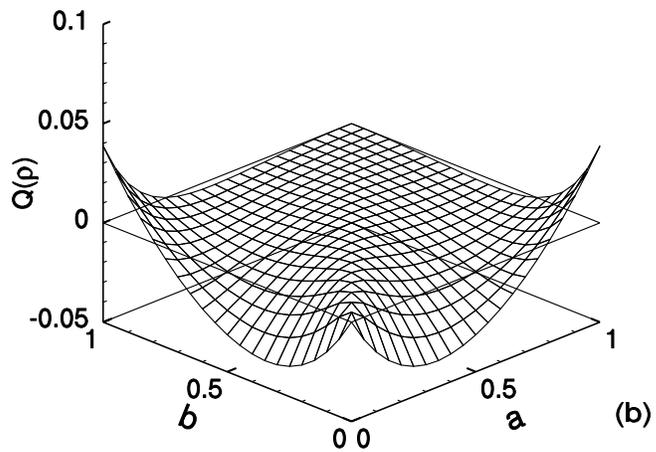
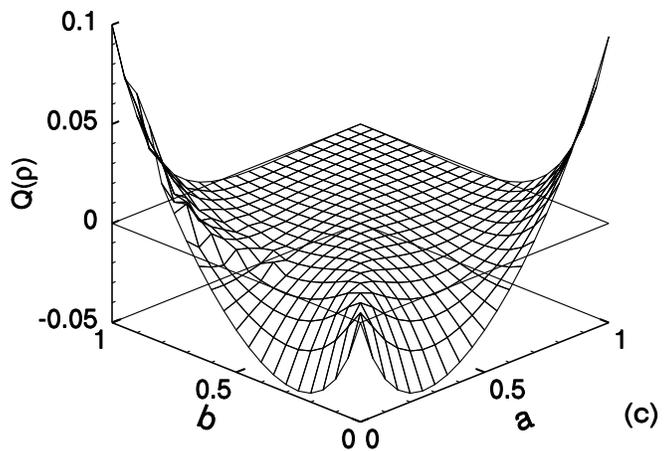
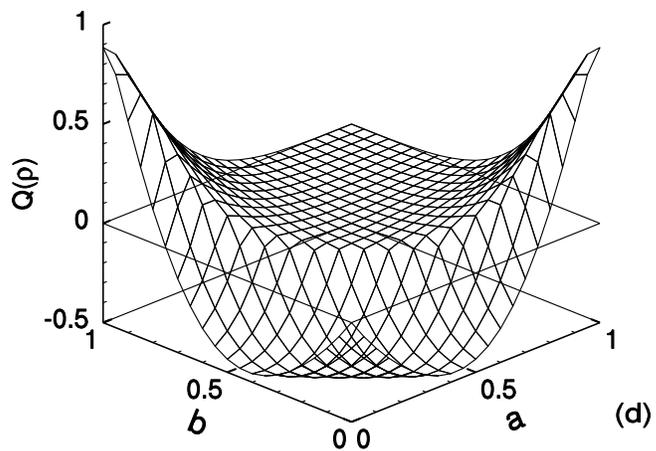
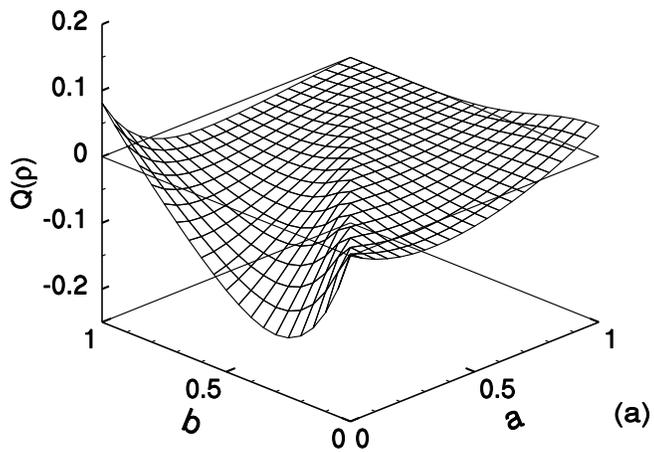
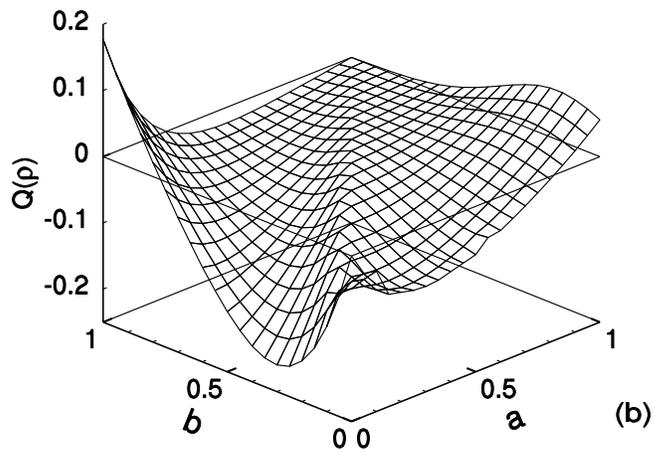
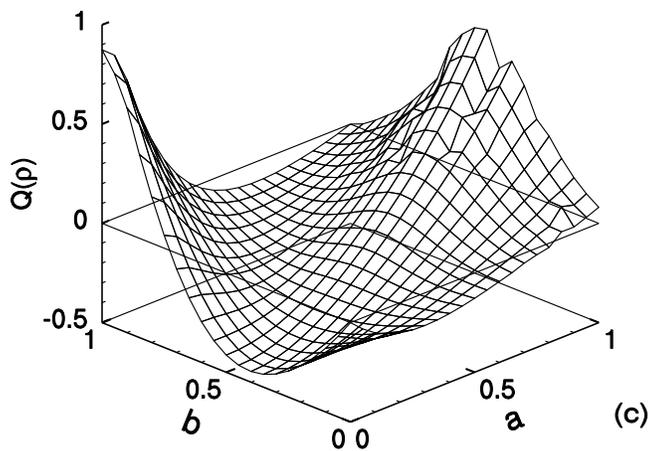
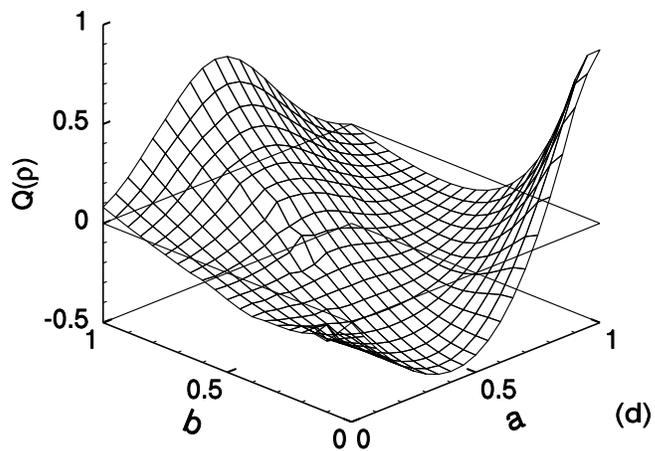
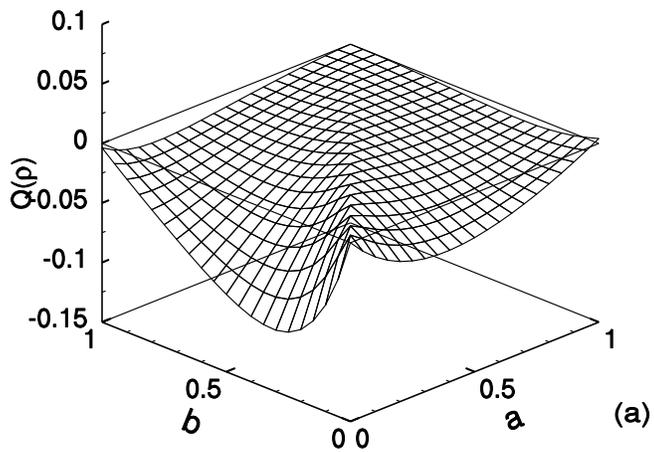
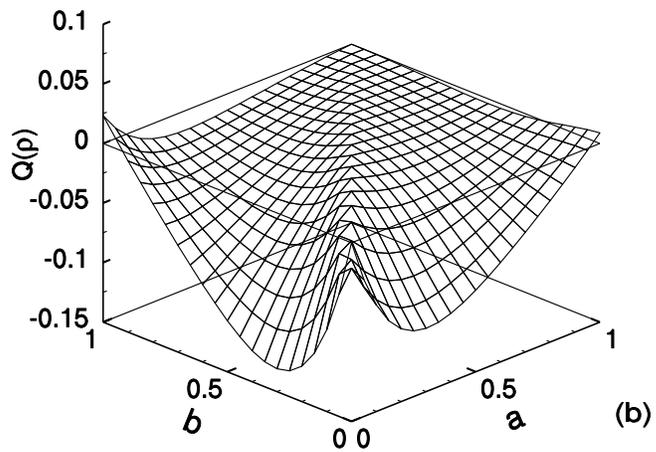
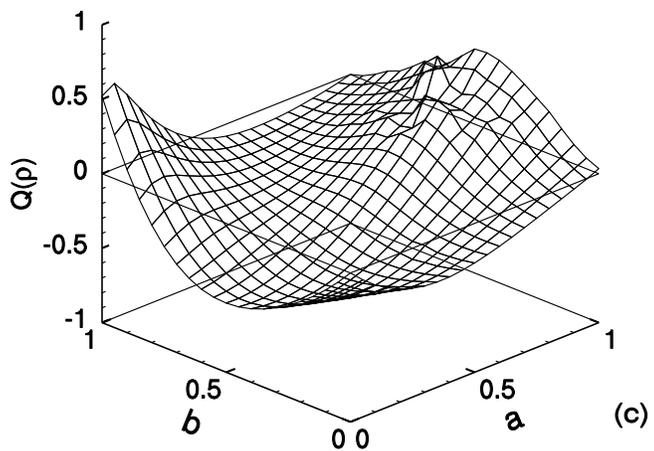
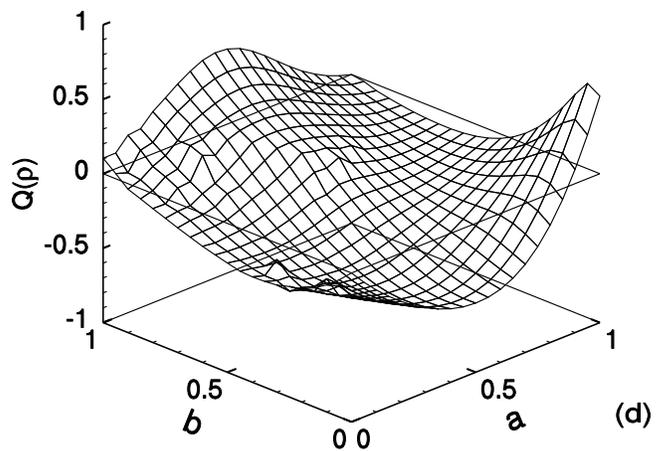
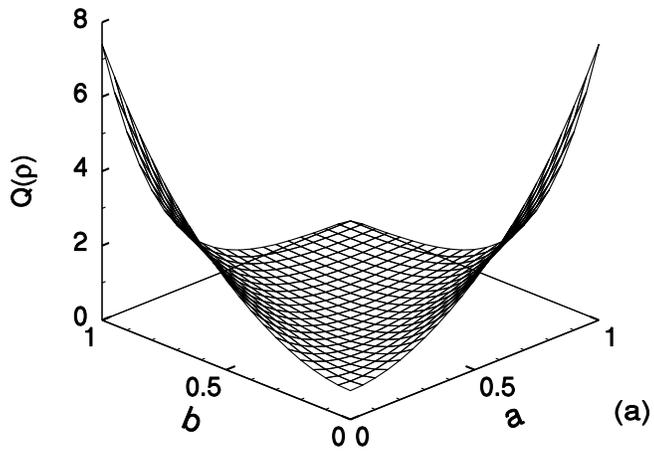
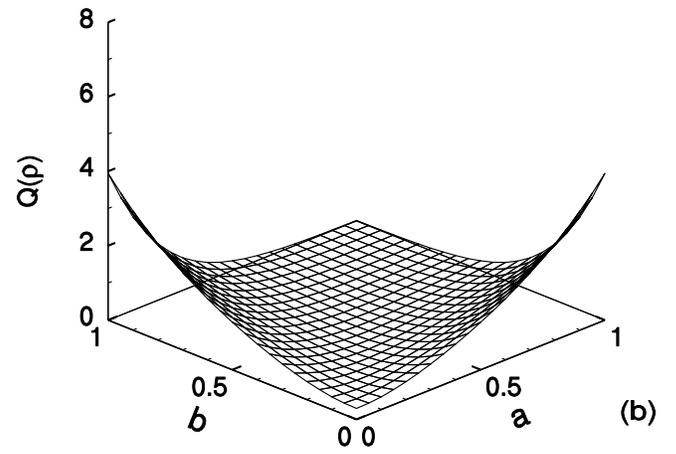
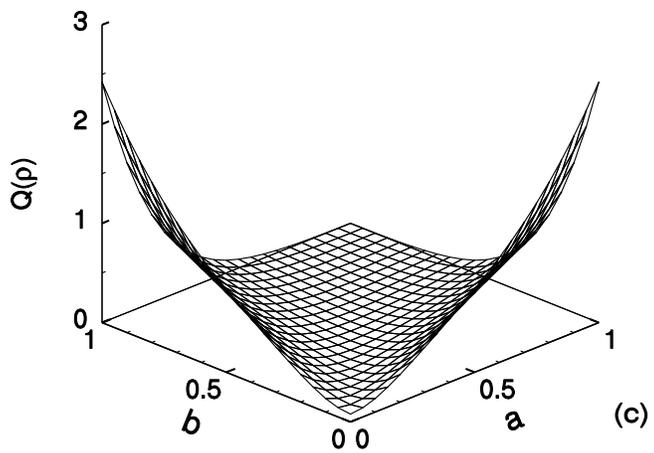
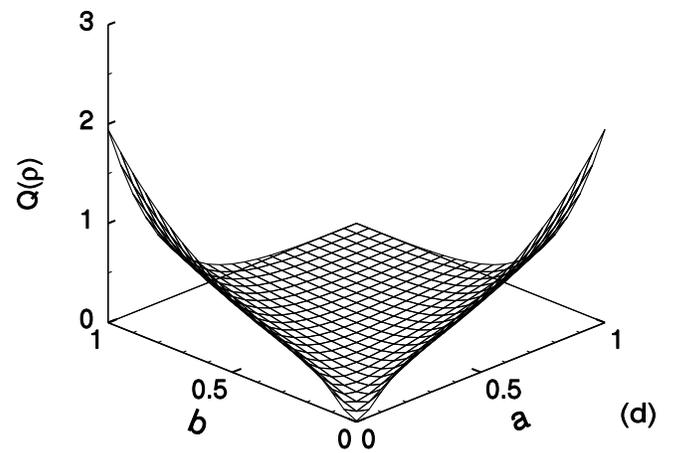
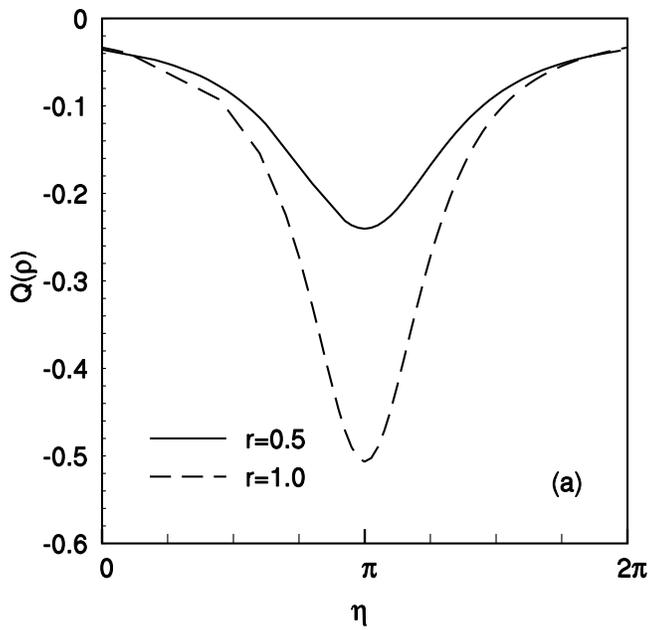
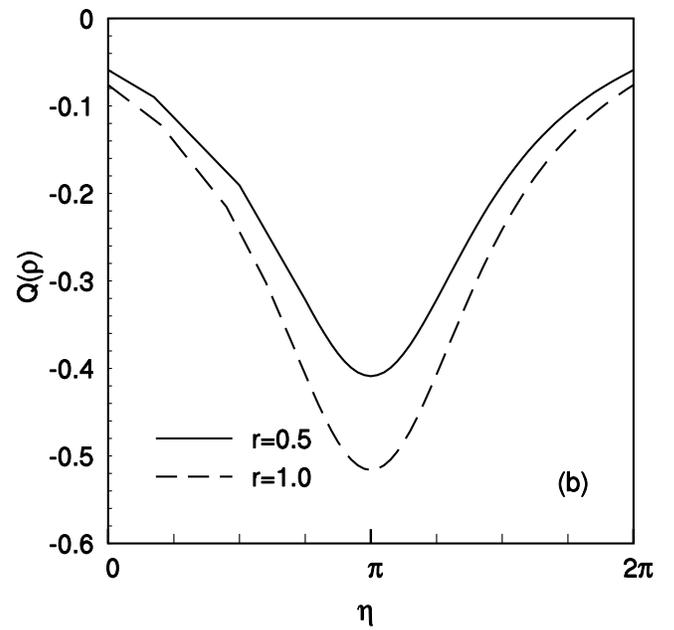
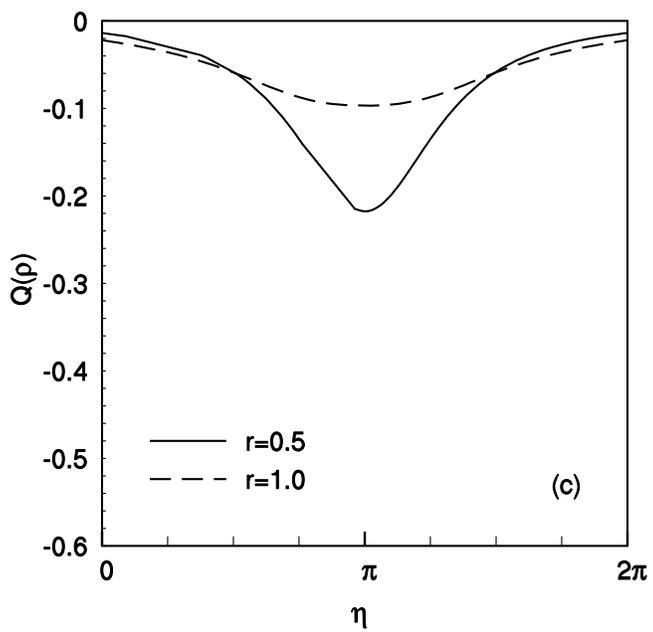
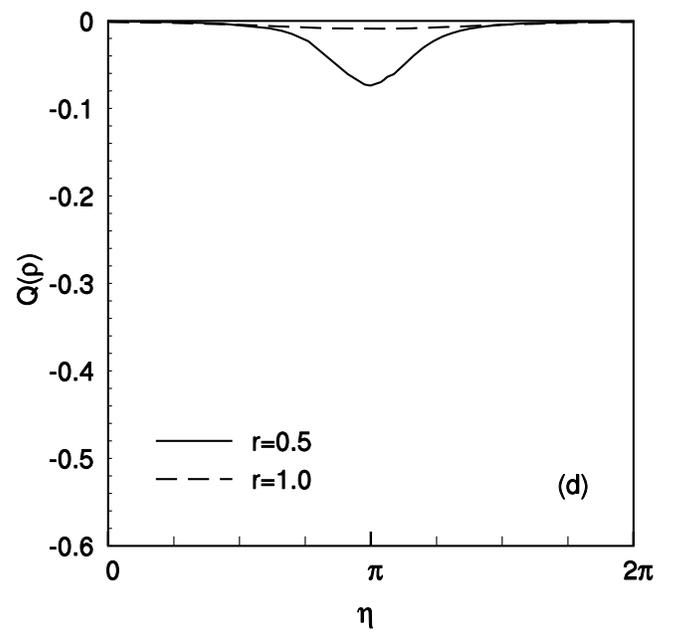
\end{document}